\documentclass{article}
\usepackage{fortschritte}
\usepackage{amsfonts}
\usepackage{amssymb,amsmath}

\newcommand{\mbf}[1]{{\boldsymbol {#1} }}
\newcommand{\complex}{{\mathbb C}} 
\newcommand{\zed}{{\mathbb Z}} 
\newcommand{\nat}{{\mathbb N}} 
\newcommand{\real}{{\mathbb R}} 
\newcommand{\rat}{{\mathbb Q}} 
\newcommand{\torus}{{\mathbb T}}
\def\e{{\,\rm e}\,}

\newcommand{\vq}{{\mbf q}}

\newcommand{\vnu}{{\mbf\nu}}

\def\ii{{\,{\rm i}\,}}
\def\dd{{\rm d}}
\def\Li{{\rm Li}}

\newcommand{\eq}{\begin{equation}}
\newcommand{\eqend}{\end{equation}}
\newcommand{\eqa}{\begin{eqnarray}}
\newcommand{\nonueqa}{\begin{eqnarray*}}
\newcommand{\eqaend}{\end{eqnarray}}
\newcommand{\nonueqaend}{\end{eqnarray*}}

\newcommand{\bma}[1]{\begin{array}{#1}}
\newcommand{\ema}{\end{array}}
\newcommand{\bc}{\begin{center}}
\newcommand{\ec}{\end{center}}

\newif\ifold             \oldtrue

\def\e{{\,\rm e}\,}

\def\beq{\begin{equation}}
\def\eeq{\end{equation}}
\def\bea{\begin{eqnarray}}
\def\eea{\end{eqnarray}}
\def\bd{\begin{displaymath}}
\def\ed{\end{displaymath}}

\begin {document}

\begin{flushright}
\baselineskip=12pt
HWM--04--18\\
EMPG--04--08\\
hep--th/0409288\\
\hfill{ }\\
September 2004
\end{flushright}
\setcounter{footnote}{1}
\def\titleline{
%
%
%
Double Scaling String Theory of QCD in Two Dimensions\footnote{
Based on invited talk given by R.J.S. at ``Recent
Developments in String/M-Theory and Field Theory'', 37th International
Symposium Ahrenshoop on the Theory of Elementary Particles,
Berlin-Schm\"ockwitz, Germany, August 23--27 2004; to be published in
{\sl Fortschritte der Physik}.}
}
\def\email_speaker{
{\tt
%
%
R.J.Szabo@ma.hw.ac.uk            
}}
\def\authors{
%
%
%
%
%
L.~Griguolo\1ad , D.~Seminara\2ad ,           
R.J.~Szabo\3ad\sp                           
}
\def\addresses{
%
%
%
%
\1ad                                        %
Dipartimento di  Fisica, Universit\`a  di Parma,
INFN Gruppo Collegato di Parma\\
Parco Area delle Scienze 7/A, 43100 Parma, Italy\\            
\2ad                                        
Dipartimento di Fisica, Polo Scientifico Universit\`a di
  Firenze, INFN Sezione di Firenze\\
Via  G. Sansone 1, 50019 Sesto Fiorentino, Italy\\       
\3ad                                        
Department of Mathematics, Heriot-Watt University\\ Scott
  Russell Building, Riccarton,
Edinburgh EH14 4AS, U.K.\\     %
}
\def\abstracttext{
%
%
We describe a novel double scaling limit of large $N$ Yang-Mills
theory on a two-dimensional torus and its relation to the geometry of
the principal moduli spaces of holomorphic differentials.
}
\large
\makefront
\section{Is Noncommutative QCD a String Theory?\label{INYMTST}}

One of the most sought goals of modern theoretical high-energy physics
is to recast the quantum field theory of the strong interactions as a string
theory. There are various indications that this might be a more
tractable problem in the context of noncommutative QCD. While
noncommutative field theories are in general most naturally induced in
string theory, they possess many unconventional stringy properties
themselves reflecting the non-locality that their interactions
retain~\cite{NCrev}. For instance, the large $\theta$ expansion of a
noncommutative field theory organises itself into planar and
non-planar Feynman diagrams in exactly the same way that the large $N$
expansion of a multicolour field theory does. They can be represented
and analysed exactly as matrix models, indicating a potential
connection to non-critical strings. Their fundamental quanta are
electric dipoles, extended rigid rods whose lengths are proportional
to their momenta, and their interactions are thus governed by
string-like degrees of freedom. Finally, some of these theories admit
novel soliton and instanton solutions which have no counterparts in
ordinary field theory and can be naturally interpreted as D-branes.

These aspects become particularly interesting in two dimensions,
because ordinary Yang-Mills theory on a Riemann surface has a very
precise interpretation as a string theory~\cite{GT1}, while gauge
theory on the noncommutative torus $\torus_\theta^2$ has
been solved exactly~\cite{PSz1}. We introduce the
dimensionful noncommutativity parameter $\Theta=A\,\theta/2\pi$ where
$A$ is the area of the torus. When $\theta=n/N$ is a rational number,
Yang-Mills theory on $\torus_\theta^2$ is exactly equivalent to
ordinary gauge theory on the torus $\torus^2=\torus_{\theta=0}^2$ with
gauge group $U(N)$. This relationship is an example of a Morita
equivalence, and its fruitfulness lies in the fact that a generic irrational
noncommutative gauge theory can be obtained as a limit of commutative
ones by taking the limits $N,n\to\infty$ with $\theta=n/N$ fixed. This
suggests that great insight into the string representation of
noncommutative gauge theory could be gotten from that of its
commutative cousin. However, the limit that is required on the
commutative side is not the conventional planar large $N$ limit. The
Morita equivalence rescales the Yang-Mills coupling as $g^2\to g^2/N$
and the area as $A\to A/N^2$ in order to ensure that the action
functionals map into each other, so that the large $N$ limit requires
a 't~Hooft scaling and a weak-coupling limit. For example, gauge theory
on the noncommutative plane $\real^2_\Theta$ can be induced from gauge
theory on the noncommutative torus $\torus_{1/N}^2$ in the limit $N\to\infty$,
$A\to\infty$ with $\Theta=A/2\pi\,N$ fixed~\cite{GSV1,GSSz1}. This
is a ``double scaling limit'' in which, due to above rescalings, the
parameter $\mu=g^2\Theta=N^2g^2A/2\pi$ is held fixed.

\section{Fluxon Expansion\label{FE}}

The quantum gauge theory on $\real_\Theta^2$ can be solved exactly
starting from the instanton expansion of $U(N)$ Yang-Mills theory on
$\torus^2$, whose partition function is given
by~\cite{PSz1,GSV1,Grig1}
\beq
Z=\e^{-\epsilon^{~}_{\rm F}}\,\sum_{\stackrel{\scriptstyle
\vnu\,:\,\sum_kk\,\nu_k=N}{\scriptstyle\sum_k\nu_k=\nu}}~
\prod_{k=1}^N\frac{(-1)^{(k+1)\nu_k}}{\nu_k!}\,
\left(\frac{2\pi^2N}{k^3\lambda}\right)^{\nu_k/2}\,
\sum_{\vq\in\zed^\nu}(-1)^{(N-1)\,\sum_kq_k}~\e^{-S_{\rm inst}}
\label{UNinstexp}\eeq
where $\epsilon^{~}_{\rm F}=-\frac\lambda{12}\,(N^2-1)$ is the
zero-point energy and $\lambda=g^2A\,N$ is the dimensionless 't~Hooft
coupling. This expansion reflects the fact that the semi-classical
approximation to two-dimensional Yang-Mills theory is {\it
  exact}~\cite{Witten1}. Each partition $\vnu$ of the rank $N$ corresponds to a
classical solution of the Euclidean gauge theory, i.e. an (unstable) instanton,
carrying a configuration of topological charges $\vq$. The action
\beq
S_{\rm inst}=\frac{2\pi^2N}\lambda\,\sum_{l=1}^N\frac1l\,
\sum_{j=1+\nu_1+\dots+\nu_{l-1}}^{\nu_1+\dots+\nu_l}q_j^2
\label{Sinst}\eeq
is the Yang-Mills action evaluated on a classical solution, while the
other terms in (\ref{UNinstexp}) represent the exact quantum
fluctuations about each instanton.

Applying the Morita transformations to (\ref{UNinstexp}) and taking
the double-scaling limit, one can
derive the fluxon expansion of gauge theory on the noncommutative
plane whose partition function is given by~\cite{GSV1,GSSz1}
\beq
\mathcal{Z}_\infty=\sum_{\ell=0}^\infty(-1)^\ell~
\e^{-S_{\rm fluxon}}\,\sum_{\vnu\,:\,\sum_kk\,\nu_k=\ell}~
\prod_{k=1}^\ell\frac{(-1)^{\nu_k}}{\nu_k!}\,\left(
\frac{A^2}{2\pi\,k^3g^2\Theta^3}\right)^{\nu_k/2}
\label{fluxonexp}\eeq
where $S_{\rm fluxon}=\pi\,\ell/g^2\Theta$ is the action of a fluxon
of charge $\ell$~\cite{fluxon1}, i.e. a solution of the noncommutative
Yang-Mills equations on $\real^2$ of finite action and magnetic flux $\ell$.
The
(infinite) area $A$ in (\ref{fluxonexp}) plays the role of an infrared
regularization. Note that these field configurations disappear in the
commutative case because $S_{\rm fluxon}\to\infty$ for $\ell>0$ as
$\Theta\to0$. Thus, in contrast to ordinary gauge theory on $\real^2$,
pure Yang-Mills theory on $\real_\Theta^2$ contains topologically
non-trivial field configurations and so even its partition function is
non-trivial. The fluxons can be interpreted as a system comprised of
$\ell$ unstable D-instantons inside a D-string. With this correspondence
in mind, we will explore the fate of the conventional
string representation of $U(N)$ Yang-Mills theory on $\torus^2$ in the
double scaling limit of the previous section. The crucial point is
that in the noncommutative setting there is no analog of a
strong-coupling heat kernel expansion available, and we must
understand how to extract stringy characteristics directly from the
weak-coupling instanton expansion.

\section{Gross-Taylor Series on the Torus\label{GTST}}

The {\it chiral} strong-coupling expansion of $U(N)$ gauge theory on
$\torus^2$ is a variant of the usual Migdal expansion whose
partition function is given by $Z^+=\sum_{R\in{\rm
    Rep}^+(U(N))}\e^{-\frac\lambda{2N}\,C_2(R)}$~\cite{GT1}, where the sum runs
through unitary irreducible representations $R$ of the
gauge group $U(N)$ whose corresponding Young diagrams contain $\ll N$
boxes, and $C_2(R)$ is the quadratic Casimir invariant of $R$. There is also an
anti-chiral sector, containing contributions from Young diagrams with
box numbers of order $N$, and the full (non-chiral) partition function
can be split into weakly interacting chiral and anti-chiral
sectors~\cite{GT1}. The string expansion of the free energy $F^+=\ln
Z^+$ can be derived via an asymptotic $\frac1N$ expansion
$F^+=\sum_{h\in\nat}F_h^+/N^{2h-2}$ with
$F_h^+=\lambda^{2h-2}\,\sum_{n\in\nat}\omega_h^n~\e^{-n\,\lambda}$,
where the non-negative integers $\omega_h^n$ are called (simple)
Hurwitz numbers and they count the number of $n$-sheeted holomorphic
covering maps $\pi:\Sigma_h\to\torus^2$ of the torus by a Riemann
surface of genus $h$ with $2h-2$ simple branch points. This
expansion contains a factor $\lambda^{2h-2}$ from the moduli space
integration over the positions of the branch points, as well as a
Nambu-Goto factor~$\e^{-n\,\lambda}$. The string coupling and tension
are given by $g_s=\frac1N$ and $T=\frac1{2\pi\,\alpha'}=\frac\lambda
A$. The anti-chiral expansion of the gauge theory yields the
anti-holomorphic sector of this string theory.

For genus $h\geq2$, the quantity $F_h^+$ is a quasi-modular
form of weight $6h-6$ under the modular group $PSL(2,\zed)$ of the
elliptic curve with modulus
$\tau=-\frac\lambda{2\pi\ii}$~\cite{quasi1}. In particular, it can be
expanded in terms of Eisenstein series~\cite{Rudd1}. This property
constitutes an important ingredient in the proof of the mirror
symmetry conjecture in the special case of elliptic curve target
spaces~\cite{BCOV1}. In addition, the quasi-modular transformation
rules for the Eisenstein series under $\tau\to-1/\tau$ allow one to analyse the
Douglas-Kazakov type singularity of the theory~\cite{DK1} in the weak coupling
limit $\lambda\to0$, at which point a condensation of instantons in
the vacuum occurs. The weak-coupling expansion was argued
in~\cite{Rudd1} to assume the generic form
\beq
F_h^+=\lambda^{2h-2}\,\sum_{k=3h-3}^{4h-3}\frac{r_{k,h}~\pi^{2(k-3h+3)}}
{\lambda^k}
\label{genformweak}\eeq
with $r_{k,h}\in\rat$. The singularity at $\lambda=0$ is
relatively mild and not the most general one expected based on the
quasi-modular behaviour. The expansion (\ref{genformweak}) is derived
by rather brute force conformal field theory techniques, and the
precise meaning of the rational numbers $r_{k,h}$ is not clear. In
what follows we will describe how to systematically derive this expansion
directly from the instanton representation of the gauge theory and
elucidate the geometric meaning of the $r_{k,h}$.

\section{Saddle-Point Expansion\label{SPE}}

Starting from the instanton expansion (\ref{UNinstexp},\ref{Sinst}) in
the limit $N\to\infty$, we will approximate the series by truncating
it to the zero-instanton sector of magnetic charges $\vq=\mbf0$. Then
one can straightforwardly derive the contour integral
representation~\cite{GSS1}
\beq
Z_0=\e^{-\epsilon_{\rm F}^{~}}\,\oint\frac{\dd z}{2\pi\ii z}~
\e^{-N\,\ln z-\sqrt{\frac{2\pi\,N}\lambda}~\Li_{3/2}(-z)}
\label{zeroinstpart}\eeq
enforcing the constraint on the sum over partitions $\vnu$ in
(\ref{UNinstexp}), where generally
$\Li_\alpha(z)=\sum_{k\in\nat}z^k/k^\alpha$ denotes the polylogarithm
function of index~$\alpha$. We evaluate (\ref{zeroinstpart}) using
saddle-point techniques. The saddle-point equation is
$\Li_{1/2}(-z)=-\sqrt{N\,\lambda/2\pi}$. Its solution can be found as
a $\frac1N$ expansion by writing $z=\e^x$ with
$x=x_{-1}\,N+\sum_{k\in\nat_0}\frac{x_k}{N^k}$ and using the
asymptotic expansion of the polylogarithm function given
by~\cite{GSSz1}
\beq
\Li_\alpha\left(-\e^x\right)=2\,\sum_{k=0}^\infty\frac{\left(1-2^{2k-1}
\right)\,B_{2k}~\pi^{2k}}{(2k)!\,\Gamma(\alpha+1-2k)}~x^{\alpha-2k}
\label{Liasymptexp}\eeq
for $x\to\infty$, where $B_{2k}\in\rat$ are the Bernoulli numbers. By
substituting (\ref{Liasymptexp}) into the saddle-point equation, one can
straightforwardly find an iterative solution in $\frac1N$ for $x$ and
evaluate the corresponding one-loop WKB approximation to
the integral (\ref{zeroinstpart}). The free energy $F_0=\ln Z_0$
agrees {\it precisely} with the form obtained in (\ref{genformweak}),
reproducing the correct rational numbers $r_{k,h}$~\cite{GSS1}. However, our
approximations have only reproduced the chiral sector of the full
gauge theory, and moreover they do not carry enough information to
reproduce the complete quasi-modular expansion of
the free energy. Part of the problem is that the integral
(\ref{zeroinstpart}) does not have a nice large $N$ scaling, so that
the saddle-point approximation is not reliable. In addition, the
collective behaviour of the higher instanton configurations, of order
$\e^{-N/\lambda}$, are crucial for the recovery of the full string
expansion.

\section{Double Scaling Limit\label{DSL}}

We can alternatively motivate the double
scaling limit, originally introduced in Section~\ref{INYMTST}, as one
in which the saddle-point analysis yields rigorous results. We take
$N\to\infty$ with the rescaled coupling
$\mu=\frac{N\,\lambda}{2\pi}=\frac{N^2g^2A}{2\pi}$ fixed, giving the
partition function
\beq
\mathcal{Z}_0=\e^{\frac{\pi\,N\,\mu}{12}}\,\oint\frac{\dd z}
{2\pi\ii z}~\e^{-N[\ln z+\frac1{\sqrt\mu}~\Li_{3/2}(-z)]}=:
\oint\frac{\dd z}{2\pi\ii z}~\e^{N\,\hat F(z)}
\label{DSLpartfn}\eeq
which admits a nice large $N$ scaling. The saddle-point equation is
now $\Li_{1/2}(-z)=-\sqrt\mu$, independent of $N$, and so it will
yield exact results in this case. It is possible to write down an
exact expression for the free energy $\mathcal{F}_0=\ln\mathcal{Z}_0$
as a primitive of the position of the saddle point $z=\e^x$ given by
\beq
\mathcal{F}_0=N\,\hat F\left(\e^x\right)=\frac N{\sqrt\mu}\,
\int_{\sqrt\mu}^\infty\dd y~\left[x\left(y^2\right)-\mbox{$\frac\pi4$}\,
y^2\right] \ .
\label{DSLfreenexact}\eeq
To understand the physical significance of the double-scaling limit,
we consider the strong-coupling limit $\mu\to\infty$ wherein we
can expect to make contact with the string picture of
Section~\ref{GTST}. The saddle-point position is now naturally an
expansion in the double-scaling parameter as
$x=\pi\,\sum_{k\in\nat_0}\xi_{2k-1}/\mu^{2k-1}$~\cite{GSS1}, and from
(\ref{DSLfreenexact}) the free energy is given by
\beq
\mathcal{F}_0=\pi\,N\,\sum_{k=1}^\infty\frac{\xi_{2k-1}}{4k-3}~
\frac1{\mu^{2k-1}} \ .
\label{freenstrong}\eeq
Comparing this with the leading contribution of the conformal field
theory expansion (\ref{genformweak}) in the double-scaling limit,
which is given by $F_0=\pi\,N\,\sum_{h\in\nat}r_{4h-3,h}/\mu^{2h-1}$,
one finds $r_{4h-3,h}=\xi_{2h-1}/(4h-3)$ and thus the double-scaling
limit resums the most singular terms in the weak-coupling limit
$\lambda\to0$ of the original chiral Gross-Taylor string
expansion of~QCD$_2$.

\section{Asymptotics of Hurwitz Numbers\label{AHN}}

The singularities of the free energy as
$\lambda\to0$ are controlled by a power-law growth of the Hurwitz
numbers $\omega_h^n$ for large degree covers. Thus,
in the double scaling limit, only strings of infinite winding number
contribute to the string expansion with
$\omega_h^n\simeq\beta_h\,n^{\alpha_h}$ for $n\to\infty$, where
$\alpha_h>0$ and by definition
\beq
\beta_h=(\alpha_h+1)\,\lim_{N\to\infty}\,\frac1{N^{\alpha_h+1}}\,
\sum_{n=1}^N\omega_h^n \ .
\label{betahdef}\eeq
Substitution of this asymptotic behaviour into the original chiral
Gross-Taylor series gives $F^+\simeq\sum_{h\in\nat}(\frac\lambda
N)^{2h-2}\,\beta_h~\Li_{-\alpha_h}(\e^{-\lambda})$ in the limit
$\lambda\to0$. By using the leading singular behaviour of the
polylogarithm function $\Li_\alpha(z)\simeq\Gamma(1-\alpha)(-\ln
z)^{\alpha-1}$ near its branch point $z\to1^-$~\cite{GSSz1}, and matching with
the
strong-coupling expansion~(\ref{freenstrong}), we can determine the
coefficients $\alpha_h,\beta_h$ of the asymptotic growth
explicitly and draw our first main conclusion: {\it The double
  scaling limit of the gauge theory free energy is the generating
  function for the asymptotic Hurwitz numbers with}
\beq
\omega_h^n~\stackrel{n\to\infty}{\simeq}~\mbox{$\frac{\pi^{2h}}
{(4h-3)!}$}~\xi_{2h-1}~n^{4h-4} \ .
\label{omegaasympt}\eeq
Thus the explicit knowledge of the small area behaviour of the gauge
theory allows one to solve the combinatorial problem of determining
the asymptotic forms of the simple Hurwitz numbers. One can
write down the exact saddle-point solution at strong coupling which
computes these numbers by setting $x=1/w^2$ in the saddle-point equation
and using the asymptotic expansion (\ref{Liasymptexp}) to reduce the
solution to the formal inversion of a Taylor series. Using an
appropriate version of the Lagrange-Burmann inversion formula to write
the explicit expansion of $x$ as a power series in $1/\sqrt\mu$, after
some combinatorial calculations one can arrive at an explicit formula
for the coefficients $\xi_{2h-1}$ of the strong-coupling saddle-point
expansion as polynomials in Bernoulli numbers, thereby determining
the exact asymptotics (\ref{omegaasympt})~\cite{GSS1}. The
resulting formula agrees {\it exactly} with that obtained
in~\cite{eskin1} from rather involved combinatorial techniques, but
our saddle-point method provides a much simpler and efficient way of
extracting the asymptotics of simple Hurwitz numbers.

\section{Moduli Spaces of Holomorphic Differentials\label{MSHD}}

We are now ready to formulate our main
geometric characterization of the rational numbers $r_{k,h}$ appearing
in (\ref{genformweak}) and of the double scaling limit. Let
$\pi:\Sigma_h\to\torus^2$ be a covering with simple ramification over
distinct points $z_1,\dots,z_{2h-2}\in\torus^2$, and denote by $\dd z$
the canonical holomorphic one-form on the torus $\torus^2$. Then the
pull-back $\dd u=\pi^*(\dd z)$ is a holomorphic one-form on the
Riemann surface $\Sigma_h$ with exactly $2h-2$ distinct simple zeroes
$u_i=\pi^{-1}(z_i)$. Conversely, given an abelian differential $\dd u$
with $2h-2$ simple zeroes on a Riemann surface $\Sigma_h$ of genus
$h$, the map $z=\pi(u)=\int^u\dd u~~{\rm mod}~\zed^2$ defines a simple
branched cover of the torus. Therefore, there is a one-to-one
correspondence between simple branched covers of $\torus^2$, and hence of terms
in the chiral Gross-Taylor series, and pairs $(\Sigma_h,\dd u)$. The
collection of isomorphism classes of pairs $(\Sigma_h,\dd u)$, with
$\dd u$ a holomorphic one-form with $2h-2$ simple zeroes on a Riemann
surface $\Sigma_h$ of genus $h$, is called a (principal) moduli space
$\mathcal{M}_h$ of holomorphic differentials~\cite{eskin1,Kont1}.

We can coordinatize this moduli space by defining a local chart
$\phi:\mathcal{M}_h\to\complex^{4h-3}$ through the period map
$\phi(\Sigma_h,\dd u)=(\,\int_{\gamma_1}\dd
u,\dots,\int_{\gamma_{4h-3}}\dd u)$, where
$\gamma_1,\dots,\gamma_{2h}$ form a canonical basis of homology
one-cycles for $\Sigma_h$ while $\gamma_{2h+i}$, $i=1,\dots,2h-3$
connect the zeroes $u_{i+1}$ to $u_1$ (These contours generate the
relative homology group
$H_1(\Sigma_h,\{u_i\};\zed)\cong\zed^{4h-3}$). This makes
$\mathcal{M}_h$ a complex orbifold of dimension
$\dim\mathcal{M}_h=4h-3$. We can use the pull-back of the Lebesgue
measure on $\complex^{4h-3}$ under $\phi$ to compute the hypervolume
of the subspace $\mathcal{M}_h'\subset\mathcal{M}_h^{~}$ defined by
the ``normalized'' pairs $(\Sigma_h,\dd u)$ of area
$\frac\ii2\,\int_{\Sigma_h}\dd u\wedge\overline{\dd u}=1$. The
computation parallels the counting of lattice points $\zed^{2(4h-3)}$
inside subsets of $\real^{2(4h-3)}$ and proceeds by introducing the
sequence of cones ${\rm
  C}_N\phi(\mathcal{M}_h'):=\{t\,\phi(\mathcal{M}_h')~|~0\leq
t\leq\sqrt N\,\}$ to define the hypervolume using the Lebesgue measure
on $\complex^{4h-3}$ as
\beq
{\rm vol}\left(\mathcal{M}_h'\right):={\rm vol}_{\complex^{4h-3}}
\bigl({\rm C}_1\phi(\mathcal{M}_h')\bigr)=\lim_{N\to\infty}\,
\mbox{$\frac1{N^{4h-3}}$}\,\left|{\rm C}_N\phi(\mathcal{M}_h')~\cap~
\left(\zed^{2(4h-3)}+\mbf b\right)\right| \ ,
\label{voldef}\eeq
where the vector $\mbf b\in\complex^{4h-3}$ is inserted in conjunction
with the quantization properties of the period map
$\phi$~\cite{GSS1}. Each point of the intersection in (\ref{voldef})
corresponds to a holomorphic covering map over $\torus^2$ of degree
$\leq N$. By further comparing with (\ref{betahdef}) and
(\ref{omegaasympt}) we arrive at the formula
\beq
{\rm vol}\left(\mathcal{M}_h'\right)=\lim_{N\to\infty}\,\frac1
{N^{4h-3}}\,\sum_{n=1}^N\omega_h^n=\frac{\xi_{2h-1}~\pi^{2h}}
{(4h-3)\,(4h-3)!} \ ,
\label{volfinal}\eeq
and we have thereby found our desired geometric characterization: {\it
  The double scaling limit of the gauge theory free energy is the
  generating function for the volumes of the moduli spaces of
  holomorphic differentials with}
\beq
\mathcal{F}_0=N\,\sum_{h=1}^\infty\frac{(4h-3)!}{(\pi\,\mu)^{2h-1}}~
{\rm vol}\left(\mathcal{M}_h'\right) \ .
\label{volgenfn}\eeq

\section{Outlook\label{Outlook}}

The string picture of the double scaling gauge
theory replaces the counting of holomorphic maps
$\pi:\Sigma_h\to\torus^2$ by the counting of holomorphic differentials
on $\Sigma_h$. It seems to correspond to a
thermodynamic limit of a free fermion theory, leading to an
infinite-volume system with finite particle density, in contrast to the
free fermion formulation of the standard 't~Hooft limit in
which the spatial density of fermions becomes infinite on a compact
space corresponding to the conformal field theory limit described by a $c=1$
theory. Among other things, the chirality and the
thermodynamic nature of the double-scaling limit are most akin to the
fluxon expansion of noncommutative gauge theory
on~$\real^2$. The less singular terms $r_{k,h}$, $k<4h-3$ in the
weak-coupling expansion (\ref{genformweak}) can be computed with some
work using the technique outlined in Section~\ref{AHN}, and they could
be related to other aspects, such as intersection indices, of the
moduli spaces $\mathcal{M}_h$. This could significantly sharpen the
mathematical understanding of the moduli spaces of holomorphic
differentials, which play an important role in ergodic theory but
whose geometry is not very well understood at present.

\bigskip

\noindent
{\bf Acknowledgements:} \ \  R.J.S. would like to thank the organisors
of the Symposium for an enjoyable and stimulating atmosphere. The work
of R.J.S. was supported in part by an Advanced Fellowship from the
Particle Physics and Astronomy Research Council~(U.K.).

\end{document}